\documentclass[english]{article}
\pdfoutput=1
\usepackage{mathptmx}
\usepackage{helvet}
\usepackage{courier}
\usepackage[T1]{fontenc}
\usepackage{units}
\usepackage{url}
\usepackage{amsmath}
\usepackage{graphicx}
\usepackage{amssymb}

\makeatletter

\usepackage[draft]{hyperref}
\hypersetup{pdfauthor={Matthew Brand},pdftitle={Specular holography}}
\makeatother

\usepackage{babel}

\begin{document}

\title{Specular holography}

\author{Matthew Brand}
\date{30 November 2010}
\maketitle

\begin{abstract}
By tooling an spot-illuminated surface to control the flow of specular
glints under motion, one can produce holographic view-dependent imagery.
This paper presents the differential equation that governs the shape
of the specular surfaces, and illustrates how solutions can be constructed
for different kinds of motion, lighting, host surface geometries,
and fabrication constraints, leading to some novel forms of holography.
\end{abstract}

\section{Holography via specular glints}

Holography, broadly understood, is any means of making a light field
that contains a range of views of a virtual 3D scene. Using only
geometric optics, it is possible to construct a sparse holographic
light field by tooling a 2D surface to control the bundle of rays it
directionally reflect or refracts. The author has been exhibiting such
``specular holograms'' in public art venues for a number of years.
This paper introduces the differential geometry that determines the
geometry of the optical surfaces, and develops solutions for several
viewing geometries.

The idea of specular holography predates wavefront holography by many
years. The centuries-old arts of metal intaglio and engine turning
exploit the fact that sharp glints on smooth convex specular surfaces
appear---via stereopsis---to float below the surface. Beginning in the
1930's, several authors have analyzed this effect
\cite{Kirkpatrick54,Lott63,Walker89,PlummerGardner92} and proposed to
use it for view-dependent imagery \cite{Weil34} and holography
\cite{Garfield81,Beaty95,Beaty03,EicherDunkelGoncalves03,AugierSanchez10,ReggEtAl2010}
using surface scratches or parabolic reflectors. All give reasonable
approximations for small ranges of viewpoints; this paper gives
the exact geometry for all viewpoints.

Fig.~\ref{thefig}(a) diagrams the optical principle: Given a point
light source and a virtual 3D point in the holographic scene, one
shapes a smooth optical surface that reflects or refracts a specular
glint to the eye along every unoccluded sightline through the point.
The surface appears dark to other sightines.  Lacking any other
depth cues, the brain parsimoniously but incorrectly concludes that
both eyes see a single specularity located at the virtual
point. Changes in viewpoint reinforce the 3D percept via motion
parallax.

A specular hologram combines a large number of such optical surfaces,
all designed to conform to a 2D host surface so that they can be easily
formed by conventional fabrication techniques such as milling, grinding,
stamping, ablation, etc. The host surface has limited surface area,
so the 3D virtual scene is represented by a sparse sampling of its
points, usually a stippling of its surfaces. As an artistic matter,
there are many ways to algorithmically produce visually informative
and pleasing stipplings. Similarly, the optical surfaces have a simple
differential geometry, developed below, that admits a variety of distinct
solutions.

\section{Geometry of a single-point optical surface}

Consider making a specular hologram of a single point
$\mathbf{p}\in\mathcal{R}^{3}$.  Let $\mathbf{i}\in\mathcal{R}^{3}$ be
a point illumination source, and let $\mathbf{e}\in\mathcal{R}^{3}$ be
the location of the eye.  Typically the viewpoint $\mathbf{e}$ is
parameterized by azimuth $\theta$ and elevation $\phi$; $\mathbf{p}$
and $\mathbf{i}$ can vary parametrically as well. Throughout this
paper, a boldfaced variable will refer interchangeably to a
three-vector and the function giving its locus. The holographic effect
is achieved by producing a specular glint on each sightline through
virtual point $\mathbf{p}$ at a real point
$\mathbf{s}\in\mathcal{R}^{3}$ where the axis of reflection or
refraction is parallel to the optical surface normal
$\mathbf{n}\in\mathcal{R}^{3}$. The ideal optical surface is a
continuous locus of points swept by $\mathbf{s}$ as
$\mathbf{p},\mathbf{i},\mathbf{e}$ vary. Generally, there is an
infinite \emph{foliation} of such surfaces, each a different distance
from $\mathbf{p}$ on the sightline. To fabricate the hologram, a host
surface $\mathbf{H}$ with local normal $\mathbf{N}\in\mathcal{R}^{3}$
is tooled to conform piecewise to this foliation, changing its local
normal to $\mathbf{n}$.

As in computational mirror design \cite{Hicks05}, one needs to
reconstruct a surface from a field of arbitrarily scaled normals.
Here it is more convenient work with tangent spaces than with
normals. Let
$T_{\mathbf{s}}=(\mathbf{t}_{1},\mathbf{t}_{2})\in\mathcal{R}^{3\times2}$
be any nondeficient basis of the optical surface's local tangent plane
at $\mathbf{s}$. For example, $T_{\mathbf{s}}$ could be the Jacobian
$J_{\mathbf{s}}(\theta,\phi)$ of $\mathbf{s}$ with respect to
$\theta,\phi$.  Then the constraints that determine the optical
surface are \begin{eqnarray}
  \left\langle T_{\mathbf{s}},\frac{\mathbf{i}-\mathbf{s}}{\|\mathbf{i}-\mathbf{s}\|_{2}}\eta_{1}+\eta_{2}\frac{\mathbf{e}-\mathbf{s}}{\|\mathbf{e}-\mathbf{s}\|_{2}}\right\rangle  & = & (0,0)\qquad\textnormal{ (normality)}\label{eq:normality}\\
  \left\langle (\mathbf{e}-\mathbf{p})^{\bot},(\mathbf{s}-\mathbf{p})\right\rangle  & = & (0,0)\qquad\textnormal{ (colinearity)}\label{eq:colinearity}\\
  \|\mathbf{s}-\mathbf{H}\| & \leq &
  \Delta\qquad\qquad\!\!\textnormal{(conformity)}\label{eq:conformance}\end{eqnarray}
where $\eta_{1},\eta_{2}$ are the refractive indices of materials a
light ray crosses before and after encountering the optical surface;
$\left\langle \cdot,\cdot\right\rangle $ and $\|\cdot\|$ are the
Euclidean inner product and vector norm; and $\mathbf{x}^{\bot}$ is
any orthogonal basis of the nullspace of $\mathbf{x}$, i.e.,
$\left\langle \mathbf{x}^{\bot},\mathbf{x}\right\rangle =0$.  For
reflection holography, $\eta_{1}=\eta_{2}\neq0$. Differential
Eq.~\eqref{eq:normality} (normality) states that the optical surface
is perpendicular to the axis of reflection or
refraction. Eq.~\eqref{eq:colinearity} (colinearity) ensures that the
specularity is on the sightline to $\mathbf{p}.$
Eq.~\eqref{eq:conformance} (conformance) specifies that the optical
surface lie within a thin shell of thickness $2\Delta$ that conforms
to the host surface $\mathbf{H}$ . The colinearity constraint can be
algebraically eliminated by sliding the eye forward along the
sightline $\mathbf{e}\mathbf{p}$ until it coincides with $\mathbf{p}$
or $\mathbf{s}$.

It is easily shown (by substitution) that Eq.~\eqref{eq:normality} is
satisfied by a foliation of revolute bicircular quartics with foci at
$\mathbf{i}$ and $\mathbf{p}$.  The quartics are revolved around the
major axis connecting $\mathbf{i}$ and $\mathbf{p}$.

\begin{figure}
\begin{centering}
\includegraphics[width=5in]{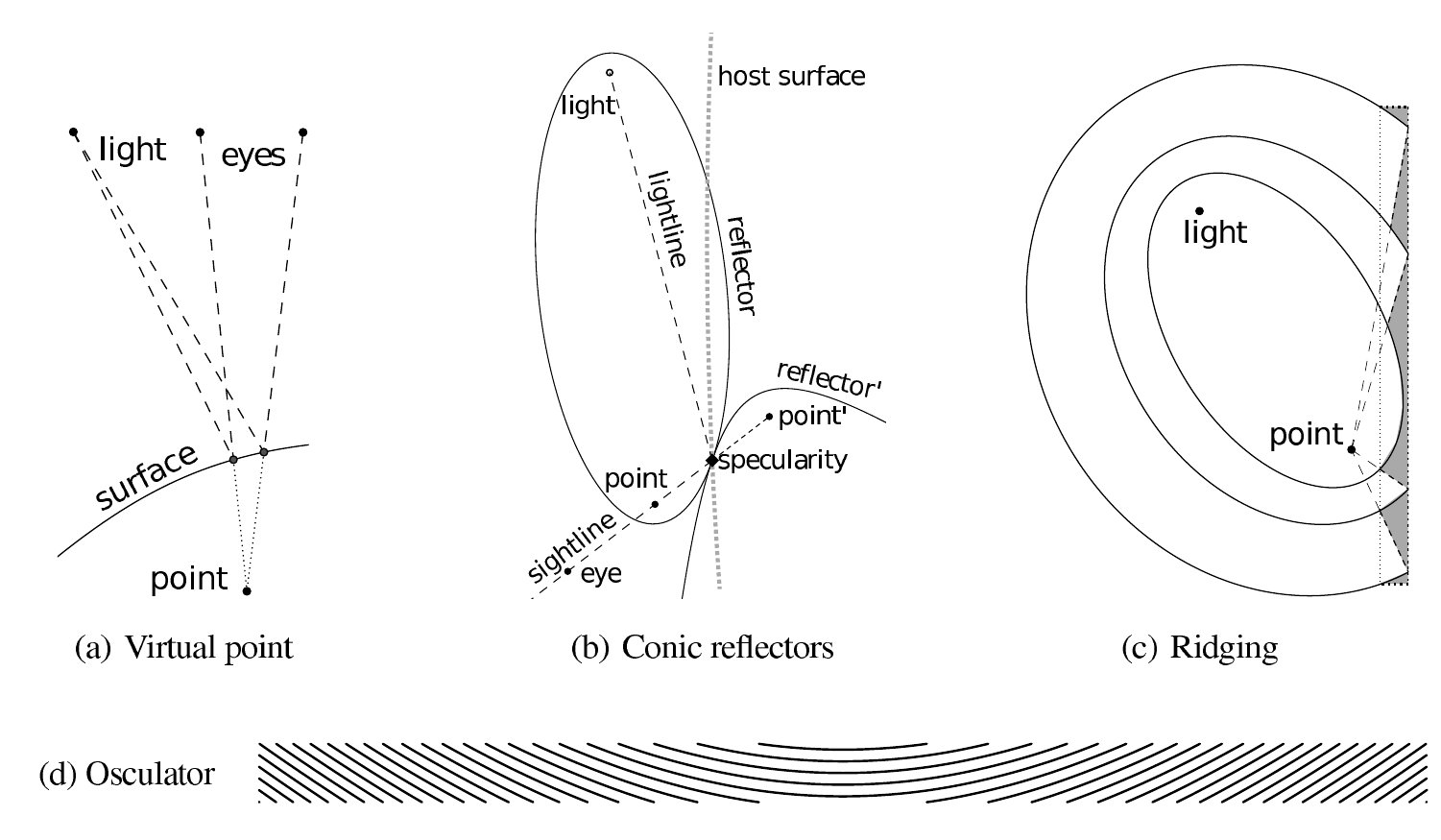}
\end{centering}
\caption{\label{thefig} Geometry of specular holography: (a) Glints on
  a curved surface are perceived as an off-surface point. (b) The
  ideal surface is any of a foliation of quartics (for reflection,
  conics) with one focus at the light and the other at the point. (c)
  An equivalent ridged surface is bounded by surfaces from the
  foliation, the host surface, and a set of cones. (d) A striping of
  toolpaths on a host surface whose swept volume osculates the
  foliation.}
\end{figure}

The most common case of interest is a reflection hologram with static
point light source $\mathbf{i}$ and virtual point $\mathbf{p}$.  In
this case the quartics simplify to conics.  Specifically, if $\mathbf{p}$
is in front of the reflecting surface, the solution is the interior
surface of any prolate ellipsoidal reflector with eccentricity
$\epsilon<1$. If point $\mathbf{p}$ is behind the reflecting surface,
the solution is the exterior surface of any hyperboloid of
eccentricity $\epsilon>1$. Fig.~\ref{thefig}(b) illustrates how these
two solutions\footnote{There are also two degenerate solutions: The
  exterior of an infinitesimal parabolic needle for $\mathbf{p}$ at
  the surface $\mathbf{s}$ ($\epsilon=1$), the interior of a sphere
  for $\mathbf{p}$ at light source $\mathbf{i}$ ($\epsilon\to0$).  }
are related.

The remaining constraint (Eq.~\eqref{eq:conformance}) can be satisfied
by constructing a ridged surface from the foliation that conforms
to the host surface. Fig.~\ref{thefig}(c) illustrates the construction.
Each ridge is the intersection of a thin shell at the host surface,
a cone emanating from some point in front of the host surface, and
a revolute conic (or its complement) taken from the foliation. This
construction works for arbitrary host surfaces, however, as with all
Fresnel-like surfaces, the resulting optical surface may self-occlude
some rays. 

In a hologram, each stipple generates a ridged optical surface, which
is cropped to a small region on the host surface to limit
azimuthal and elevation visibility intervals. Cropping can be
used to suggest occlusion, illumination, and shading effects in the
virtual scene; and for viewpoint-keyed animation.

An early instance of an approximate specular holographic surface is
the 1961 VITA solar cooking stove. In this case, the sun is at
$\mathbf{i}$ (effectively, infinity), the pot is at $\mathbf{p}$, and
the host surface is a plane perpendicular to the ellipsoid's medial
axis. A conic with one focus at infinity is a paraboloid, hence the
stove uses a Fresnel paraboloidal reflector.  Stoves with finer
ridgings produce strong depth illusions.  A similar construction is
used in \cite{ReggEtAl2010}, who report producing a depth illusion for
an azimuthal view range of up to $30^\circ$, albeit with distortions.

For refracting holograms, the exact optical surfaces are generally not
conics.  For single-interface optical paths, Eq.~\eqref{eq:normality}
is solved by a foliation of revolute Cartesian ovals.  Proper modeling
of two-surface optical paths typically leads to numerical integration,
with some notable exceptions. Given a planar back surface, a focusing
front surface can be derived from Fermat's principle, yielding an
parametric sag function.  In typical settings the point light source
is far enough behind the back surface that the ray bundle incident on
the front surface is approximately radially divergent, in which case
the front surface can be very tightly approximated by a revolute
Cartesian oval with one focus shifted from $\mathbf{i}$ to the center
of divergence.  This approximation becomes exact for divergent light
on a spherically-backed surface and for collimated light on a
flat-backed surface, in which case the classical geometry of
plano-aspheric lenses (e.g, \cite{KweonKim07}) provides a foliation of
hyperbolic or elliptic front surfaces, depending on the ratio of
refractive indices $\eta_1/\eta_2$ and whether the virtual point
$\mathbf{p}$ is in front of or behind the refracting surface.  Of
course, if both front and back surfaces are machined, the problem
reduces to classic aspheric lens design.

\section{Horizontal parallax via osculating surfaces}

Ridged surfaces are imperfect optics: Some rays founder on the non-imaging
backface of each ridge, causing stray specularities and dead spots.
In practice, ridged surfaces proved to be tedious and difficult to
machine cleanly at small scales. This motivated a switch to smoother
osculating surfaces that are easy to fabricate rapidly and finely.
An osculating surface {}``kisses'' the foliation by matching its
surface normals on a bounded submanifold of contact points. I.e.,
there is some region where the osculating surface produces the same
specularities. A trivial example arises when machining an optical
surface with a ball-end milling bit: Each pass of the bit sweeps a
volume whose surface osculates the desired surface. Swept volumes
can be machined much faster and smaller than ridged surfaces, enabling
far more detailed stippings in less time. For example, a recently
exhibited hologram has roughly $10^{5}$ stipples; current fabrication
methods are approaching $4$ stipples/second for $90^\circ$-view holograms. 

A swept volume that provides horizontal parallax to all viewpoints can
be determined analytically for many viewing
geometries: Eq.~\eqref{eq:normality} determines a tangent space for
every point in $\mathcal{R}^{3}\backslash\{\mathbf{p},\mathbf{i},\mathbf{e}\}$.
Integrating the component of these tangents that conforms to the host
surface yields a foliation of toolpaths that sweep the
surface. Integrating the orthogonal component yields the profile of
the cutting bit.

To illustrate, consider a reflective optical surface embedded in a
locally flat host surface, lit by the sun at azimuth $0$, elevation
$\pi/2-\alpha$ (at high noon, $\alpha=0$), and producing a light
field with horizontal parallax. Here it is useful to imagine an infinitely
thin and long mirrored cylinder laid flush on the host surface and
angled such that an eye looking along a ray from azimuth $\theta$, elevation
$0$ sees a specular glint. Using a local coordinate frame centered
on the specularity, the cylinder axis is a vector through the origin
in the $z=0$ plane, the eye is at $\mathbf{e}\propto(\sin\theta,0,\cos\theta),$
and the sun is at $\mathbf{i}=\infty\cdot(0,\cos\alpha,\sin\alpha)$.
The local host surface normal is $\mathbf{N}=(0,0,1)$ and the optical
surface normal $\mathbf{n}$ at the glint is \[
\mathbf{n}\propto\mathbf{i}\|\mathbf{e}\|+\mathbf{e}\|\mathbf{i}\|\propto(\sin\theta,\cos\alpha,\cos\theta+\sin\alpha)\:.\]
As per Eq.~\eqref{eq:normality}, the local tangent space $T_{\mathbf{s}}=(\mathbf{t}_{1},\mathbf{t}_{2})$
of the optical surface at glint $\mathbf{s}$ must be orthogonal to
$\mathbf{n}$; i.e., $\mathbf{n}\propto\mathbf{t}_{1}\times\mathbf{t}_{2}$.
Without loss of generality, choose optical surface tangent $\mathbf{t}_{1}$
to lie in the local tangent space of the host surface: \begin{equation}
\mathbf{t}_{1}\propto\mathbf{n}\times\mathbf{N}\propto(\cos\alpha,-\sin\theta,0)\:.\end{equation}
Writing $\mathbf{t}_{1}\propto(dx,dy,dz)$ yields the slope of the
cylinder axis: $\frac{dy}{dx}=-\frac{\sin\theta}{\cos\alpha}$. 

Integrating this slope w.r.t. view angle $\theta$ yields the toolpath
$y(\theta)$ that satisfies Eq.~\eqref{eq:normality} as the viewpoint
revolves around a virtual point $\mathbf{p}=(0,0,p_{z})$. To this
end, one needs the fact that the sightline from $\mathbf{e}$ to $\mathbf{p}$
passes through a flat host surface at horizontal location $x(\theta)=-p_{z}\tan\theta$.
Integrating then reveals that the toolpath is a hyperbola: \begin{eqnarray}
y(\theta)=\int\frac{dy}{dx}\frac{dx}{d\theta}d\theta & = & -p_{z}\sec\alpha\sec\theta+C_{0}\textnormal{ ,}\label{eq:toolpath}\\
\textnormal{thus  }\: y(x) & = & p_{z}(\sec\alpha)[(x/p_{z})^{2}+1]^{\nicefrac{1}{2}}+C_{0}\:.\label{eq:hyperbola}\end{eqnarray}
The horizontal locations of specularities on this toolpath are parallax-consistent,
but because the specularities travel up and down the {}``arms''
of the hyperbola, they are not on the sightline and therefore Eq.~\eqref{eq:colinearity}
is not satisfied. This can be solved by noting that the integration
constant $C_{0}$ specifies a foliation of identical hyperbolas, all
differing by vertical offsets. We take the intersection of this foliation
with a thin rectangular bar, then select a subset of discrete hyperbolic
arcs that can be machined without destructive overlap. That produces
a\emph{ striping} (Fig.~\ref{thefig}(d)) similar to the
fringe pattern of a single-point hogel in a Benton white-light hologram.

The toolpath sweeps through the conforming tangents of the optical
surface.  The orthogonal tangents \begin{equation}
  \mathbf{t}_{2}\propto\mathbf{t}_{1}\times\mathbf{n}\propto(-\sin\theta,-\cos\alpha,\frac{\cos^{2}\alpha+\sin^{2}\theta}{\cos\theta+\sin\alpha})\label{eq:OrthTangent}\end{equation}
can be integrated to yield a space curve that osculates the cutting
bit, from which the bit profile can be deduced. Because any positive
integration measure can be used for $d\theta$, it suffices that the
bit profile merely has the range of surface-to-tangent angles
exhibited by $\mathbf{t}_{2}$ along the toolpath. For example,
assuming an overhead light ($\alpha=0)$ and a flat host surface, the
angle $\angle\mathbf{N},\mathbf{t}_{2}$ varies by $15^{\circ}$ over a
$90^{\circ}$ range of azimuthal viewpoints. Consequently the profile
of the cutting bit needs to be slightly curved to produce a swept
volume with all the required normals. This curvature also serves to
make the holographic images visible to eyes at positive and negative
elevations, albeit without vertical parallax.

To extend the swept volume construction to a curved host surface with
normal function $\mathbf{N}(\theta,\phi)$, assume that the eye moves
along some path (e.g., a transverse line), producing a specularity
space curve where sightlines through the virtual point intersect the
host surface. Define $x(\theta)$ on this curve and calculate $y(\theta)$
and$ $ $z(\theta)=\int\frac{dz}{dx}\frac{dx}{d\theta}d\theta$ w.r.t.
$x(\theta)$ and $\mathbf{N}(\theta,\phi)$ to obtain a foliation
of toolpaths $(x(\theta),y(\theta)+C_{0},z(\theta)+C_{1})$. The desired
striping is the intersection of this foliation with a thin shell around
the specularity space curve. Generally these curves are not conics.

By appropriate parameterization, the toolpath-and-profile construction
will also yield stripings for holograms with arbitrarily located point
light sources, discontinuous host surfaces, and motions of the virtual
points, light source, or host surface. 

Note that the unstriped hyperbolic toolpath is well approximated by a
circular arc for a roughly $8^{\circ}$ range of azimuthal
viewpoints---sufficient for stereopsis from a central viewpoint. This
explains the appeal of scratch
holography\cite{Weil34,Garfield81,Beaty95,Beaty03,EicherDunkelGoncalves03,AugierSanchez10},
and also why its holographic image distorts and collapses outside that
range.

The ease of fabrication of long toolpaths motivates another class of
solutions---to be presented in a follow-up paper---where
Eq.~\eqref{eq:normality} is used to solve for motions of the virtual scene or
the host surface such that all specularities on the \emph{continuous}
(unstriped) toolpath provide rigidly rotated views of the scene.

Roughly 100 holograms have been fabricated over the last few years
using all three classes of solutions (ridgings, stripings, and
continuous toolpaths), including unusual variants such as animated
holograms, continuous $720^{\circ}$-view\footnote{For example, a
  hologram in which the view circles a 3D head twice, seeing a
  different face in each cycle.}  planar holograms, and holograms on
curved surfaces. From early 2009, pieces have been exhibited in
public art venues \cite{NOS09,A10}, attracting large crowds.  As with
all holograms, the pieces are difficult to photograph satisfactorily;
so readers are invited to see videos at
\small{\url{http://www.zintaglio.com}}~.

\end{document}